\documentstyle[pre,aps,epsf,multicol]{revtex}
\begin{document}
\newcommand{\be}{\begin{equation}}
\newcommand{\ee}{\end{equation}}

\title{One-dimensional Nonequilibrium Kinetic Ising Models 
with local spin-symmetry\\
breaking: N-component branching annihilating random-walk
transition at zero branching rate}
\author{N\'ora Menyh\'ard}
\address{Research Institute for Solid State Physics and Optics, 
\\ H-1525 Budapest,P.O.Box 49, Hungary}
\author{G\'eza \'Odor}
\address{Research Institute for Technical Physics and Materials Science, 
\\ H-1525 Budapest, P.O.Box 49, Hungary}
\maketitle

\begin{abstract}
The effects of locally broken spin symmetry are investigated in
one dimensional nonequilibrium kinetic Ising systems
via computer simulations and cluster mean field calculations.
Besides a line of directed percolation transitions,
a line of transitions belonging to N-component, two-offspring branching 
annihilating random-walk class (N-BARW2) is revealed in the phase diagram 
at zero branching rate.
In this way a spin model for N-BARW2 transitions is proposed for the 
first time.
\end{abstract}


\begin{multicols}{2}

\section{ Introduction }

The Ising model is a well known static, equilibrium model. Its
dynamical generalizations, the kinetic Ising models, were originally 
intended to study relaxational processes near equilibrium states 
\cite{gla63,kaw72}. Glauber introduced the single spin-flip kinetic 
Ising model, while Kawasaki  constructed a spin-exchange version for 
studying the case of conserved magnetization.
Nonequilibrium kinetic Ising models, in which the steady
state is produced by kinetic processes in connection with heat
baths at different temperatures, have been widely investigated
and results have shown that various phase transitions are possible under 
nonequilibrium conditions, even in one dimension (1d)
(for a review see the article by R\'acz in Ref. \cite{rac96}).
Most of these studies, however, have been concerned with the effects
the nonequilibrium nature of the dynamics might exert on phase
transitions {\it driven by  temperature}.

A different line of investigating nonequilibrium
phase transitions has been via branching annihilating random 
walk (BARW)  processes.
The parity conservation of particles is decisive
in determining the universality class of the phase transition.
A coherent picture of this scenario is provided from a renormalization
point of view in \cite{Card97}.
The first example of a BARW model with an even number of 
offsprings exhibiting the so called PC (parity conserving) 
transition was reported by Grassberger et al.\cite{gra8489}.

A class of general nonequilibrium kinetic Ising models
(NEKIM) with combined spin flip dynamics at $T=0$ and Kawasaki spin
exchange dynamics at $T=\infty$ has been proposed by one of
the authors \cite{men94} in which, 
for a range of parameters of the model,  PC-type transition
takes place. This model has turned out to be  very rich in several
respects, for a review see\cite{Braz}.
 
Absorbing transitions have been, however mostly studied in particle-type
models. The N-BARW2 model is a classical stochastic system of $N$ types
of particles with branching annihilating random walk and two offsprings. 
For $N=1$ the model exhibits, at finite branching rate $p$, 
PC type transition [8-14].
For $N>1$ $N$ types of particles $A_i$ perform diffusion, pairwise
annihilation of the same species and branching $A_i \rightarrow A_i+2A_j$
with rate $p$ for $i=j$ and with rate $p^,/{(N-1)}$ for $i\neq j$.
In case of $p=0$ this model is always active except for the 
annihilation fixed point at zero branching rate. According to field 
theory \cite{Card97} the coarse-grained, bosonic version of the model 
forms a different universality class, the so called N-BARW2 with exponents 
in one dimension as follows:
$\nu_{\bot}=1$, $z=2$, $\alpha=1/2$, $\beta=1$
Here the exponents are defined as follows.:
\begin{equation}
\xi \sim p^{-\nu_{\bot}} , \tau\sim \xi^z
\label{ksi}
\end{equation}
\begin{equation}
\rho(t) \sim t^{-\alpha}, \rho_{\infty}\sim p^\beta
\label{rho}
\end{equation}
$\tau$ is the characteristic time, $\xi$ is the correlation length and 
$\rho(t)$, $\rho_{\infty}$ are the particle densities at time $t$ and in 
the steady state, respectively.

Hard core interactions have proven to be relevant in case of the N-BARW2
model by drastically changing the universality class \cite{kwonetal,barw2cikk}.
The arrangement of the offsprings relative to the parent turns out
to be a relevant factor and causes two robust classes that are insensitive
to parity conservation \cite{dp2cikk} or the binary nature of the production
process \cite{parwcikk}.

In this paper we  present an asymmetric spin-model (NEKIMA) with
asymmetry both in the annihilation and spin-flip rate as a generalization
of NEKIM. On the level of kinks, however, this model corresponds
to a process of A and B particles with $A \rightarrow ABA$ and 
$B\rightarrow BAB$ -type branching and $AB \rightarrow 0$ annihilation 
and $BA \rightarrow BA$ exclusion.
Nevertheless, as will be presented below using computer simulations,
the critical properties near the zero branching limit are the same
as for the N-BARW2 model cited above with no sign of exclusion effects
since alternating sequences of $A$-s and $B$-s occur like in \cite{HCV01},
hence hard core interactions can not play an important role.
Moreover, at finite branching rate of the kinks a line of
directed percolation (DP) -type transition \cite{DP} occurs
which is well described by $N=6$ level cluster mean-field calculations.
\section{The model}

The general form of the Glauber spin-flip transition rate in
one-dimension for spin $s_i$ sitting at site $i$  is \cite{gla63} 
($s_i=\pm1$):
\begin{equation}
 w(s_i,s_{i-1},s_{i+1})={\Gamma\over{2}}(1+\delta s_{i-1}s_{i+1})\left[
(1 - {1\over2}s_i(s_{i-1} + s_{i+1})\right]
\label{Gla}
\end{equation}
at zero temperature. Usually the Glauber model is understood as
the special case  $\delta=0$, $\Gamma=1$.

The Kawasaki spin-exchange transition rate of neighbouring spins \cite{kaw72} 
at $T=\infty$ reduces to an unconditional nearest neighbour exchange:
\begin{equation}
w_{ex}(s_i,s_{i+1})={ p_{ex}\over{2}}(1-s_{i}s_{i+1})
\label{Kaw}
\end{equation}
where $p_{ex}$ is the probability of spin exchange.
The quantities in Eqs. (\ref{Gla}) and (\ref{Kaw}) conserve
spin symmetry, of course.
Concerning  spin exchanges, which  act only at domain  boundaries,
the process of main importance here is that a kink can produce two
offsprings at the next time step with probability
\begin{equation}
p_{k\rightarrow3k}\propto{p_{ex}}.
\label{pex}
\end{equation}
By changing $p_{ex}$ for {\it negative values of $\delta$} this model
displays phase transitions in the  parity conserving 
universality class \cite{men94}.

In the following we will be interested in investigating an extended
version of the above model. Instead of (\ref{Gla}) we will prescribe the rates
for the case $\Gamma=1$, $\delta=0$ as follows. The $+++ \rightarrow +++$ and 
$--- \rightarrow ---$ processes remain as in (\ref{Gla}) i.e. at zero 
temperature no kink-pair creations occur inside of domains.
Further rates will be chosen in such a way that they break the symmetry of 
$+$ and $-$ spins locally.  
Such dynamically self-induced field was first investigated in a different
context by Majumdar, Dean and Grassberger (MDG) \cite{mdg}, namely in
studying the $T=0$ coarsening dynamics of an Ising chain in a local
field which favors $-$ spins as compared to $+$ ones dynamically. 

In addition to the choice in \cite{mdg} concerning the asymmetry in the
annihilation rate
\begin{eqnarray}
w(+;--)=1  \\
w(-;++)=0
\end{eqnarray}
further  spin symmetry breaking will be introduced here, namely,
in the spin-flip part of the Glauber transition rate
the strength of which  will be measured by a further parameter $p_{+}$.
While the transition rate
\begin{equation}
w(-;+-)=w(-;-+)=1/2
\end{equation}
is unchanged,
the two rates flipping $+$ spins will be reduced as
\begin{equation}
w(+;+-)=w(+;-+)=p_{+} < 1/2
\label{p+}
\end{equation}
in order to balance the effect of the other dynamically induced field
arising from (6) and (7) by locally favoring $+$ spins.
The spin-exchange part of the model remains as in the the spin-symmetric case,
(\ref{Kaw}).

In the terminology of domain walls or particles the following
reaction-diffusion picture arises. There are two kinds of domain walls: 
$-+\equiv A$ and $+-\equiv B$ which can only occur alternately because of the
spin background. Upon meeting $AB \rightarrow 0$ while in the opposite
sequence, $BA$, the two domain-walls are repulsive due to (7). 

The absorbing states in the extreme situation $p_{+}=0$ when spin flipping
maximally favors $+$ spins, are states with single frozen $-$ spins like 
$+-+++-++-+++$. By increasing $p_{+}$, a slow random walk of these 
lonely $-$- spins starts and by annihilating random walk only one of them 
survives and performs RW. The all $+$ and all $-$ states are,
of course, also absorbing.

The inclusion of nearest neighbor spin exchange, (\ref{Kaw}), changes the 
picture drastically.
Spin exchange leads to $A\rightarrow ABA$ and $B\rightarrow BAB$ type kink
production, which, together with $AB \rightarrow 0$ annihilation
($BA \rightarrow 0$ is forbidden due to the annihilation-asymmetry, 
eqs.(6) and (7)) and diffusion of $A$ and $B$ leads to a kind of 
two-component, coupled branching and annihilating random walk.
The phase diagram and the nature of the transitions will 
be reported and discussed in the following.

\section{phase diagram of  the NEKIMA model}
\subsection{The line of DP transitions}
\subsubsection{Simulation results}

Given three independent parameters $\delta$, $p_{+}$ and $p_{ex}$ (or
rather $p_{ex}/\Gamma$ ) with the restriction eq. (\ref{p+})
it is hard to explore the whole phase diagram.
In the original (spin-symmetric) case of NEKIM \cite{men94} we have 
investigated the phase boundary in the parameter space $(\delta, p_{ex})$.
For {\it negative} values of $\delta$ a line of PC transitions was found. 
We are not going to investigate the $\delta < 0$ case in the following and 
only make the remark that spin-asymmetry as introduced above 
(with a trivial generalization for $\delta <0$), changes the parity 
conserving character of the transitions to directed percolation 
- type (of two species), as could be expected. 
The case $\delta \geq 0$ (including the Glauber case $\Gamma =1, \delta=0$)
was found to be Ising-like, for all values of $p_{ex}$.

\begin{figure}[h]
\epsfxsize=80mm
\epsffile{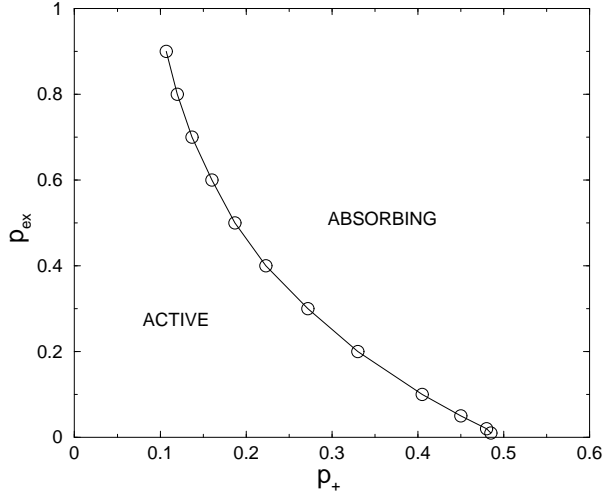}
\caption{Phase diagram of the NEKIMA model for $\delta=0$, $\Gamma=1$.
The absorbing (fully $-$ ) phase lies above the boundary. The $p_{+}=0.5$
point is referred to as MDG point in the text }
\label{phasedia}
\end{figure}

Introducing spin-asymmetry, however, makes the Glauber case also richer
in phases. The phase diagram in the plane of parameters $p_{ex}$ and
$p_{+}< 1/2$ as obtained by computer simulations is shown on
Fig.\ref {phasedia}. 
In fixing the phase boundary the quantity measured was the
density of kinks, $\rho(t)$, as a function of time starting from a random
initial distribution of up and down spins. The chain size $L$ varied between
$L=2000-5000$ and up to $t=5\times 10^5$ Monte Carlo steps (MCS) were 
reached. The way of updating was as in ref.\cite{men94}. A line of DP 
transitions has been found by the power-law behaviour of 
$\rho(t) \sim t^{-\alpha}$
with $\alpha= .160 \pm .005$, the value characteristic of DP transition.
This was, of course, the expected kind of order-disorder transition
on the basis of spin-asymmetry.

The point $p_{+}=1/2$, $p_{ex}=0$ is a particular one, at this
point our model goes over to the one investigated by Majumdar 
et al. \cite{mdg}.
For these parameter values a very high precision computer measurement
has given the result $\rho(t) =t^{-1/2}/ln({t/\tau})$ supporting an analytic
independent interval approximation by these authors.
In  an equivalent model, however, the deviation from the  $t^{-1/2}$
law has shown up as an initial density-dependent power function with
a power of $\rho(t)$ slightly deviating from $1/2$ \cite{odme2000}.
This problem, however, is not the subject of the present investigation.

Concerning the phase diagram, Fig. (\ref{phasedia}), the parameter values
in the vicinity of $p_{+}=1/2$ for $p_{ex} \neq 0$ are hard from
the computational point of view as long transients show up in the 
time evolution. It is apparent, however, that the phase line ends up
here tangentially. 
At  $p_{+}=1/2$ the effect of the exchange term is such that
 for all $p_{ex}>0$ the absorbing phase is entered: due to the choice in 
eqs.(6) and (7), 
the all $s_i=-1$ phase (one of the absorbing phases) is 
reached exponentially fast.

As to  the other limiting situation $p_{+}=0$, for $p_{ex}=0$
the initial spin distribution freezes in. As a matter of
fact, the line of phase transitions reaches the $p_{+}=0$ axis
only by letting $p_{ex}/\Gamma \rightarrow \infty$ by $\Gamma 
\rightarrow 0$ ( $\Gamma$
is only fixing the rate of flips, see eq.(3), while here we 
fixed it to unity). This circumstance, however,
is of no importance for the results.

\subsubsection{Cluster mean-field calculations for the phase diagram}

Cluster mean-field approximation introduced for nonequilibrium models by
\cite{gut87,dic88} was applied for the present model.
The $N=1$ mean-field equation for spin-up density is
\begin{equation}
\frac{\partial\rho_1}{\partial t} = - 2 p_{+} (1-\rho_1)\rho_1^2
\end{equation}
independently form $p_{ex}$, which gives a $\rho_1\propto t^{-1}$
leading order decay to the $\rho_1(\infty) = 0$ solution for all 
$p_{+}>=0$,
while it is constant (keeps the initial value) for $p_{+}=0$.
Therefore this predicts a discontinuous transition along the $p_{+}=0$
axis.
The corresponding steady state exponent is $\beta=0$. Similarly the
kink density, $\rho_1(1-\rho_1$), decays with a leading order singularity
$\rho_{kink}\propto t^{-1}$ and exhibits a jump at the $p_{in}=0$ axis.

The $N=2$ pair approximation results in the following steady state
solution for kinks
\begin{equation}
\rho_{kink}(\infty) = \frac{4\,p_{ex}\,\left( 1 - 2\,p_{+} \right) \, 
p_{+}}
{1+2\,\left(4\,p_{ex}-1\right)\,p_{+} + 8\,p_{ex}\,\left(2\,p_{ex}-1\right
)\,p_{+}^2}
\end{equation}
\begin{center}
\begin{figure}[h]
\epsfxsize=80mm
\epsffile{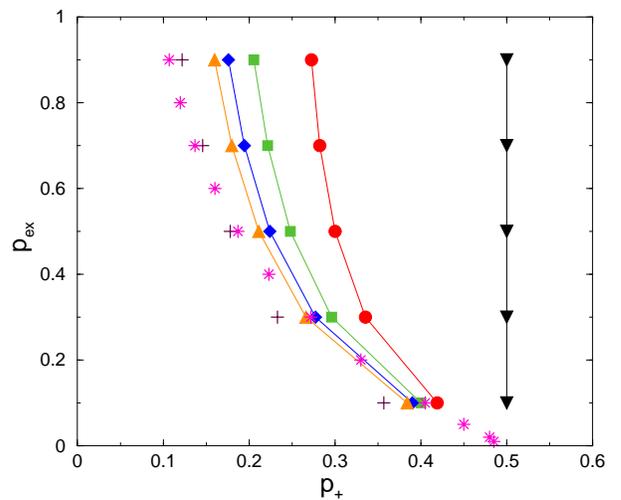}
\vspace{1mm}
\caption{Phase diagram determined by $N=2...6$ cluster mean-field 
approximations (filled symbols from right to left), the $N\to\infty$ 
extrapolated values (plus signs) and simulation results (stars).}
\label{GMFphd}
\end{figure}
\end{center}
This has absorbing states ( $\rho_{kink}=0$) along the $p{+}=0$, 
$p_{ex}$=0 and $p_{+}= 1/2$ lines and active in the $0<p_{+}<1/2$, 
$p_{ex}>0$ region. The transitions, however are continuous with leading 
order singularity $\beta=1$ everywhere. As we can see the simple mean-field 
and higher order cluster mean-field approximations give different singular 
behavior similarly to cases treated earlier\cite{odszo,meod95,Car,2dpcpd}.

For higher order approximations  $N>2$ we could solve the equations for
the steady state numerically only. We could determine stable solutions up to
the $N=6$ level from the coupled non-linear equations of 36 variables.
By locating the phase transition lines we found that the $p_{+}=0$ and
the $p_{ex}=0$ transitions do not change but the $p_{+}=1/2$ (DP)
transition line shifts monotonically towards the $p_{+}=0$ axis as we
increase the level of approximations (see Fig. (\ref{GMFphd}). These
solutions converge towards the phase transition line determined by
simulations. We found that fairly good quadratic fitting can be applied for 
the $N=3,4,5,6$ level $p_{+}^*(p_{ex},N)$ critical point solutions, so we
extrapolated to $N\to\infty$ at $p_{ex} = 0.1,0.3,0.5,0.7,0.9$.
The corresponding $p_{+}^*(p_{ex},\infty)$ curve agrees well with the
simulation data (see Table \ref{DPlinetab}.

\subsection{The line of N-BARW2 transitions}

As we have seen in the previous subsection, in the plane of 
$(p_{+},p_{ex})$
the phase below the phase transition line is the active one
and extends down to the $p_{ex}=0$ axis. The critical behavior at and
in the neighborhood of this axis has turned out to be of N-BARW2 type.
In this respect the absorbing states are fully ordered or consist of 
single $-$ -es performing random walk in the sea of $+$ spins.
\begin{figure}[h]
\epsfxsize=80mm
\epsffile{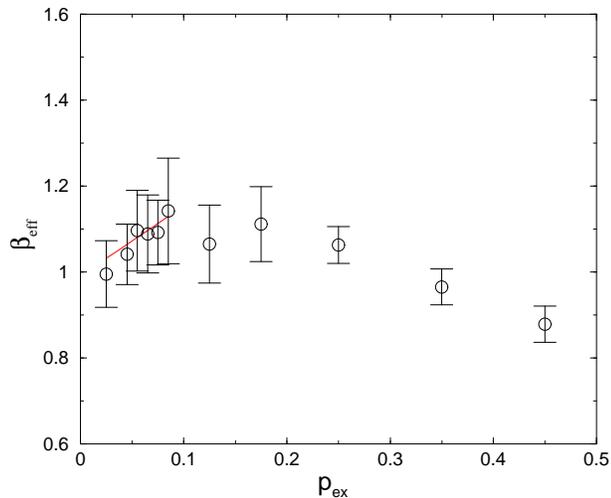}
\caption{Effective critical exponent of the order parameter at the $p_{ex}=0$ 
transition.}
\label{betaeff}
\end{figure}
The order parameter also in this case is the density of kinks
$\rho$, the steady state value of it disappears when approaching the 
$p_{ex}=0$ axis as $\rho_{\infty}\propto p_{ex}^{\beta}$. 
Simulations from random initial state 
in a system with size $L=10^5$ were run up to $10^6$ MCS.
In the supercritical region the steady states have been determined for
different $p_{ex}$ values. Following level-off the densities were 
averaged over $10^4$ MCS and $1000$ samples.
By looking at the effective exponent defined as
\begin{equation}
\beta_{eff}(p_{ex}(i)) = \frac {\ln \rho_{\infty}(p_{ex}(i)) -
\ln \rho_{\infty}(p_{ex}(i-1))} {\ln p_{ex}(i) - \ln p_{ex}(i-1)} \ \ ,
\end{equation}
one can read-off: $\beta_{eff}\to\beta$. The result of computer simulations 
at $p_+=0.1$ is shown on Fig.(\ref{betaeff}). A linear extrapolation for 
$p_{ex} < 0.1$ gives $\beta=1.0\pm .01$.
The overshooting of $\beta_{eff}$ near the critical point is typical
in case of logarithmic corrections to scaling. 
By plotting $\rho_{\infty}/p_{ex}$
as the function of $\ln(p_{ex})$ fairly good linear behavior could be
observed in the $0.02 < p_{ex} < 0.4$ region.

\begin{figure}[h]
\epsfxsize=80mm
\epsffile{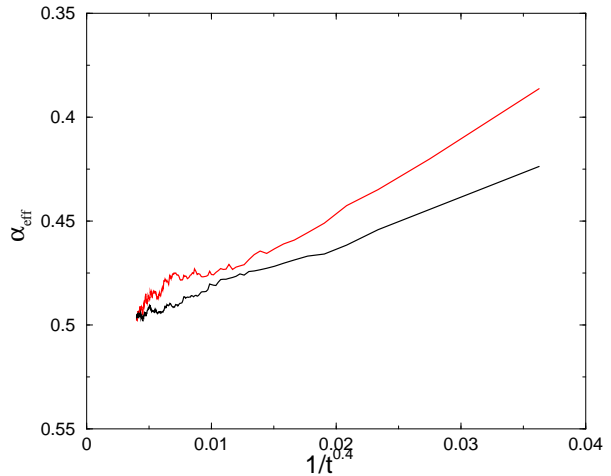}
\caption{Effective critical exponent of the kink density at the $p_{ex}=0$,
$p_{+}=0.1$ (lower curve) and $p_{+}=0.4$ (upper curve) as the function
$1/t^{0.4}$. This choice results in linear plot of the local slopes,
corresponding to $\sim t^{-0.9}$ correction to scaling.}
\label{alpha}
\end{figure}

The $\rho(t)$ simulation results at $p_{ex}=0$ and $p_{+}=0.1, 0.4$ were 
analyzed by the local slopes 
\begin{equation}
\alpha_{eff}(t) = {- \ln \left[ \rho(t) / \rho(t/m) \right] 
\over \ln(m)} \label{slopes}
\end{equation}
(where $m=8$ is used). The asymptotic time evolution of the density
of kinks $\rho(t)\sim t^{-\alpha}$ has proven to be, within error,
that of annihilating random walk: $\alpha=1/2$ as shown on fig.(\ref{alpha}).
A $\sim t^{-0.9}$ correction to scaling gave best fit in both cases. 
We also tried to fit a logarithmic correction form $((a+b\ln(t))/t)^{0.5}$
for $\rho(t)$ but $b$ was found to go zero for $t > \sim 6\times 10^5$ MCS
in both cases.

\begin{figure}
\epsfxsize=80mm
\epsffile{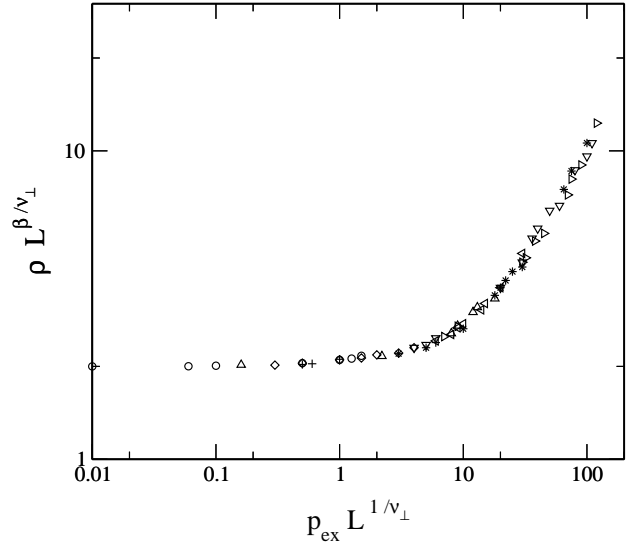}
\vspace{4mm}
\caption{Data collapse of $\rho L^{\beta/\nu_{\bot}}$ against
$p_{ex}L^{1/\nu_{\bot}}$ with  $\beta/\nu_{\bot}=1$ for various
values of the chain-lengths (circle $L=50$, diamond $L=100$,
plus sign $L=600$, triangle $L=800$, triangle left $L=1000$,
star $L=2000$, triangle right $L=3000$, triangle down $L=4000$)
on a double logarithmic scale.}
\label{scalefun}
\end{figure}

As to the  remaining critical exponents when approaching $p_{ex}=0$ we 
checked the expected N-BARW2 behaviour by measuring
the kink density in the active state (steady state) for several small
values of $p_{ex}$ (between $.1$ and $.001$) for lattice size $L$ between
$50$ and $5000$. 
The initial state was prepared in such a way that a cluster of $'-'$-es
was chosen of width and location randomly distributed between $L/4$ and $3L/4$.
Finite size scaling theory \cite{au} predicts the form
\begin{equation}
\rho(p_{ex},L)=L^{-\beta /\nu_{\bot}}F(p_{ex}L^{1/{\nu_{\bot}}})
\end{equation}  
Using the value of $\beta$ as obtained above, we determine $\nu_{\bot}$ 
by data-collapsing. With $\beta/\nu_{\bot}=1$ on Fig.(\ref{scalefun}),
we get  $\nu_{\bot}=1.00\pm .06$. Thus our result shows the critical
exponent values by Cardy and T\"auber \cite{Card97}.

\section{Summary}

In this paper a new model has been presented by generalizing
the nonequilibrium kinetic Ising model for the case when two kinds 
of spin anisotropies are present.In addition to the local kinetic
bias introduced by Majumdar et al \cite{mdg} first, by prescribing
with probability equal to zero the annihilation of $-$ spins in the
neighborhood $'+-+'$, we introduce spin-anisotropy in the spin-flip
rate. Namely, the $'+'$ spins are less likely to flip at domain 
boundaries (with probability $p_+$) than $'-'$-es (probability $1/2$).
Branching of kinks (domain boundaries) is the main effect of the
spin-exchange part of the model. We have shown by cluster mean field
calculations and computer simulations that for a given $p_{+}<1/2$ 
asymmetry the presence of spin exchange gives rise to two different 
types of phase transitions. While at $p_{ex}^*=0$ an active phase 
emerges with an N-BARW2 type of transition, for $p_{ex}^*>0$  this 
transforms back into an absorbing state with a DP class transition.
The spin anisotropies result in a new type of two-component, coupled 
branching and annihilating random walk of kinks with parity conservation.

At $p_{+}=1/2$ the MDG point is reached, see Fig. 1. It is the endpoint 
of the DP transition line similarly to the compact directed percolation 
endpoint of the DP transition line in the Domany-Kinzel cellular
automaton model \cite{DK}. The absorbing phase is the same in the two
models, the active phase, however, is different.
 
The question arises whether the  $1/\ln(t)$ factor in the
asymptotic behaviour found in \cite{mdg} is to be expected to hold
even in the present model for $p_{ex}=0$. If this were the case,
the N-BARW2 behavior would also be affected.
First, from the side of simulations, we have not found any sign of such
behavior. Moreover, the physical picture behind the expected asymptotic
behaviour of spins is also different in the two cases. While MDG argue
that at late time the process $-+- \rightarrow -$ with  probability unity
leads to $'+'$ domains  sandwiched between much larger $'-'$ ones
the introduction of  the local asymmetric spin-flip magnetic field with 
bias for $'+'$ spins will act against and feed up the $'+'$ phase to 
compensate for their biased annihilation via the MDG process.
As a consequence there is no reason to expect a late time logarithmic
relaxation of the kink density on the $p_{ex}=0$ line for $p_{+}<1/2$.

Finally it is worth  noticing that while the PC transition is known to be
sensitive to the $Z_2$ symmetry \cite{parkh,meod96} and DP transition
appears by destroying it, the N-BARW2 transition seems to be insensitive 
to this symmetry breaking.
\begin{table}
\label{DPlinetab}
\caption{Summary of $p^*_+$ critical point results of $N=3-6$ GMF
approximations and simulations}
\vspace{2mm}
\begin{tabular}{|l|c|c|c|c|c|c|}
$p_{ex}$ & $N=3$  & $N=4$ & $N=5$ & $N=6$ & $N\to\infty$ & MC \\
\hline
0.1      & 0.419  & 0.4003 & 0.3903& 0.384 & 0.357  & 0.405  \\
\hline 
0.3      & 0.3357 &0.2963 &0.2772 &0.2661 & 0.233 & 0.272   \\
\hline
0.5      & 0.3000 &0.2479 &0.2242 &0.2111 & 0.178 & 0.187   \\
\hline
0.7      & 0.2824 &0.2216 &0.1945 &0.1798 & 0.146 & 0.137  \\
\hline
0.9      & 0.2726 &0.2057 &0.1759 &0.1597 & 0.122 & 0.107 \\
\end{tabular}
\end{table}
\begin{table}
\label{NBARW2linetab}
\caption{Summary of critical exponent estimates at the $p_{ex}=0$ line.
The last row shows the data of the 1d N-BARW2 class.}
\vspace{2mm}
\begin{tabular}{|l|c|c|c|}
$p_+$    & $\beta$  & $\alpha$ & $\nu_{\perp}$ \\
\hline
0.1      & 1.00(1)  & 0.505(5) & 1.00(6) \\
\hline 
0.4      &  -       & 0.503(5) &  -  \\
\hline\hline
N-BARW2  & 1        & 1/2      &  1 \\
\end{tabular}
\end{table}

{\bf ACKNOWLEDGEMENTS} \\

The authors would like to thank the Hungarian research fund OTKA (Nos.
T034784, T017493, T025386 and 4012) for support during this study.
G. \'Odor acknowledges support from research fund Bolyai (No. BO/00142/99)
and from IKTA projekt (Project No. 00111/2000) 
The simulations were performed on the parallel cluster of SZTAKI and on the 
supercomputer of NIIF Hungary.

\end{multicols}

\begin{references}
\bibitem{gla63} R. J. Glauber,  J. Math. Phys. {\bf 4} 191 (1963).
\bibitem{kaw72} see e.g. {\it Kawasaki K:}
 Phase Transitions and Critical Phenomena,Vol.2.,\hfil
 ed.{\it Domb C and Green M S}
 (New York: Academic, 1972) p.443
\bibitem{rac96}Z. R\'acz, "Kinetic Ising models with competing dynamics:
mappings, steady states and and phase transitions" in {\it Nonequilibrium
Statistical mechanics in one Dimension} ed.V.Privman (Cambridge University
Press, Cambridge, 1996)
\bibitem{Card97} J. Cardy and U. T\"auber, 
Phys. Rev. Lett. {\bf 77}, 4780 (1996).
\bibitem{gra8489} P. Grassberger  F. Krause  and T. von der Twer
J. Phys. A:Math.Gen. {\bf 17} L105 (1984) and 
P. Grassberger, J. Phys. A:Math.Gen. {\bf 22} L1103 (1989).
\bibitem{men94} N. Menyh\'ard, J.Phys. A: Math.Gen. {\bf 27}, 6139 (1994).
\bibitem {Braz} N. Menyh\'ard, G. \'Odor,
 Brazilian J. of Physics {\bf 30}, 113 (2000).
\bibitem{taka} H. Takayasu, A. Yu Tretyakov, 
Phys. Rev. Lett. {\bf 68}, 3060 (1992).
\bibitem{je93} I. Jensen, Phys. Rev. E {\bf 47}, R1 (1993).
 \bibitem{jen94} I. Jensen, Phys.Rev.E {\bf 50}, 3623 (1994).
\bibitem{dani} D. Zhong, D. ben-Avraham, Phys. Lett. A {\bf 209}, 333 (1995).
\bibitem{kim94} M. H. Kim, H. Park, Phys. Rev. Lett. {\bf 73}, 2579 (1994).
\bibitem{park} H. Park, M. H. Kim, H. Park, Phys. Rev. E {\bf 52}, 5664 (1995).
\bibitem {parkh} H. Park and H. Park, Physica A {\bf 221}, 97 (1995).
\bibitem{DP} for a review see J.Marro and R.Dickman Nonequilibrium
phase transitions in lattice models (Cambridge University Press,
Cambridge, 1996); H. Hinrichsen Adv. Phys.{\bf 49}, 815 (2000).
\bibitem{mdg} S. N. Majumdar, D. S. Dean and P. Grassberger Phys. Rev. Lett.
{\bf 86}, 2301 (2001).
\bibitem{kwonetal} S. Kwon, J. Lee and H. Park, 
Phys. Rev. Lett.{\bf 85}, 1682 (2000). 
\bibitem{barw2cikk} G. \'Odor, Phys. Rev. E {\bf 63}, 021113 (2001).
\bibitem{dp2cikk} G. \'Odor, Phys. Rev. E {\bf 63}, 0256108 (2001).
\bibitem{parwcikk} G. \'Odor, Phys. Rev. E {\bf 65}, 026121 (2002).
\bibitem{HCV01} J. Hooyberghs, E. Carlon, C. Vanderzande,
Phys. Rev. E {\bf 64}, 036124 (2001).
\bibitem{odme2000} G.\'Odor and N.Menyh\'ard 
Phys. Rev. E {\bf 61}, 6404 (2000).
\bibitem{gut87} H. A. Gutowitz, J. D. Victor and B. W. Knight,
Physica {\bf 28D}, 18 (1987).
\bibitem{dic88} R. Dickman, Phys.Rev. {\bf A38}, 2588 (1988).
\bibitem{odszo} G. \'Odor and A. Szolnoki, 
Phys. Rev. E {\bf 53}, 2231 (1996).
\bibitem{meod95} N. Menyh\'ard and G. \'Odor, 
J. Phys. A. {\bf 28}, 4505 (1995).
\bibitem{meod96}  N. Menyh\'ard and G. \'Odor, 
J. Phys. A. {\bf 29}, 7739 (1995).
\bibitem{Car} E.~Carlon, M.~Henkel and U.~Schollw{\"o}ck,
Phys. Rev. E {\bf 63}, 036101-1 (2001).
\bibitem{2dpcpd} G. \'Odor, M. C. Marques and M. A. Santos,
{\it Phase transition behavior of a 2d PCPD model}, Phys. Rev. E in press.
\bibitem{au} T. Aukrust, D. A. Browne and I. Webman, 
Phys. Rev. A {\bf 41}, 5294 (1990).
\bibitem{DK} E. Domany, W. Kinzel, Phys. Rev. Lett.{\bf 53}, 311 (1984).

\end{references}
\end{document}